\documentclass[aps,prd,preprint,nofootinbib]{revtex4}
\usepackage{amsfonts,amsmath,amssymb}
\usepackage{array}
\usepackage{mathrsfs}
\usepackage{graphicx}
\usepackage{multirow}
\usepackage{epsfig}
\usepackage{graphicx}
\usepackage{subfigure}
\usepackage{booktabs}
\usepackage{array}
\usepackage{tabularx}
\usepackage{slashed}
\usepackage{braket}
\usepackage{bm}
\usepackage{diagbox}
\usepackage{makecell}
\usepackage{makecell}

\begin{document}
\baselineskip 20pt
\title{Radiative decay of fully-heavy tetraquark into quarkonium}
\author{\vspace{1cm} Hao Yang$^1$\footnote[1]{Contact author: yanghao2023@scu.edu.cn}, Songlin Lyu$^{2,3}$\footnote[2]{Contact author: songlin.lyu@na.infn.it}, and Bingwei Long$^{1,4}$\footnote[3]{Contact author: bingwei@scu.edu.cn}\\}

\affiliation{
\mbox{$^1$College of Physics, Sichuan University, Chengdu, Sichuan 610065, China}\\
\mbox{$^2$Dipartimento di Matematica e Fisica, Universit\`a degli Studi della} 
\mbox{Campania ``Luigi Vanvitelli'', viale Abramo Lincoln 5 - I-81100 Caserta, Italy}\\
\mbox{$^3$Istituto Nazionale di Fisica Nucleare, Complesso Universitario di Monte  S. Angelo,}
\mbox{Via Cintia - I-80126 Napoli, Italy}\\
\mbox{$^4$Southern Center for Nuclear-Science Theory (SCNT), Institute of Modern Physics,}
\mbox{Chinese Academy of Sciences, Huizhou 516000, Guangdong, China}\vspace{0.6cm}
}

\begin{abstract}
We present a comprehensive study of the radiative decays of fully heavy tetraquarks ($T_{4c}$, $T_{4b}$) and mixed heavy-flavor tetraquarks ($T_{bc\bar{b} \bar{c}}$​) into charmonium/bottomonium states, within the nonrelativistic QCD factorization framework. Numerical result indicates that the decay widths of fully charm tetraquark $T_{4c}$ into $\gamma J/\psi(\eta_c)$ are around 1 MeV. Crucially, the $\gamma J/\psi$ signal channel receives less experimental background near the $J/\psi J/\psi$ threshold, providing a compelling way to search for the possible tetraquark states X(6200). For $T_{4b}$, the decay width is highly suppressed by the bottom quark mass, just lying in tens eV level. We further find that the decay widths of $T_{bc\bar{b}\bar{c}} \to \gamma \Upsilon$ exceed those to $\gamma J/\psi$ by 3-4 orders of magnitude, indicating preferential $c\bar{c}$-pair annihilation over $b\bar{b}$. These radiative decay modes can be measured in the future experiments, and are helpful to understand the inner structure of the full heavy tetraquark.	
\end{abstract}

\maketitle

\section{INTRODUCTION}
Quantum Chromodynamics (QCD) serves as the foundational theory describing the strong interactions between quarks and gluons, which are confined within hadrons. According to the quark model, hadrons are classified into conventional categories, such as meson, baryon and exotic state, including multiquark, glueball and hybrid. The first exotic fully charm tetraquark X(6900) is observed by the LHCb Collaboration \cite{LHCb:2020bwg} through di-$J/\psi$ decay channel and promptly confirmed by the CMS and ATLAS Collaborations \cite{CMS:2023owd,ATLAS:2023bft}. Subsequently, two additional resonances near 6500 MeV and 7100 MeV are also reported. The quantum numbers of the three resonances are determined to be $J^{PC} = 2^{++}$ in the recent analyses \cite{CMS:2025fpt}.

However, to date, no fundamental principle has been established to definitively rule out specific internal structures for these tetraquark states. Studies based on various models—such as the diffusion Monte Carlo method \cite{Gordillo:2020sgc}, the extended relativized quark model \cite{Lu:2020cns} and the constituent quark model \cite{An:2022qpt}, consistently suggest a compact configuration. This is characterized by a root-mean-square radius between charm quarks of less than 0.5 fm. In contrast, a loosely bound molecular state would be expected to have a significantly larger radius, exceeding 1 fm, as predicted in the case of the $T_{cc}$ tetraquark \cite{Deng:2021gnb}. Moreover, Regge trajectory analyses \cite{Zhu:2020xni,Zhu:2024swp,Patel:2025rsf} suggest that these three states may constitute a family of related states. Simultaneously, a consistent hadronic molecule description would require the dedicated construction of two charmonia with masses near the production threshold, it appears unnatural for describing such complex three (even four) resonance structures. Moreover, the force mediated by charm meson is less likely to account for a tightly bound structure.

Following the compact tetraquark framework, a natural question arises, where to find the $0^{++}$ states, which are expected to be the ground states in the tetraquark mass spectrum. In the threshold region of 6.2–6.4 GeV, the background from double-$J/\psi$ production is extremely strong, causing the resonance information to be obscured by the background. By contrast, the $J/\psi+\gamma$ channel offers a much weaker background in this mass region, which facilitates the extraction of resonance signals. Moreover, if the mass of a resonance lies below the $J/\psi\mbox{-}J/\psi$ threshold, the $J/\psi+\gamma$ channel becomes one of the few viable reconstruction modes available. Therefore, investigating the radiative decay of fully-charmed tetraquarks into $J/\psi+\gamma$ is both essential and urgent. Moreover, this decay channel can proceed through a mechanism in which one $c\bar{c}$ pair undergoes recombination while another $c\bar{c}$ pair annihilates into a photon. This yields a significantly larger branching fraction compared to the diphoton channel, where both $c\bar{c}$ pairs must annihilate.

The existence of fully heavy constituents within the tetraquark implies the nonrelativisticity. As a result, the heavy quarks tend to remain spatially proximate and can effectively be treated as heavy–heavy diquark and anti-diquark. Accordingly, the decay and production mechanisms of such fully heavy tetraquark are well suited to description within the framework of nonrelativistic QCD (NRQCD) \cite{Bodwin:1994jh}. Followed with extensive studies toward the fully heavy tetraquark decay \cite{Sang:2023ncm,Zhang:2023ffe,Liu:2025mxv,Feng:2026orq} and production \cite{Feng:2020riv,Feng:2020qee,Huang:2021vtb,Feng:2023agq,Feng:2023ghc,Bai:2024ezn,Bai:2024flh,Wang:2025hex,Liang:2025wbt}, in this paper, we investigate the radiative decays of fully heavy tetraquark $T_{4c/b}$ and mixed heavy tetraquark $T_{bc\bar{b}\bar{c}}$ into charmonium or bottomonium.

The rest of this paper is organized as follows. In Sec. II, we present the primary formulas employed in the calculation. In Sec. III, the numerical decay widths of various fully heavy tetraquark configruations are discussed. The last section is reserved for summary and conclusions.

\section{FORMULATION}
In this paper, we consider the decay of S-wave fully heavy tetraquark into quarkonium ($J/\psi, \eta_c, \Upsilon$ and $\eta_b$) plus photon within the framework of NRQCD factorization. According to quantum chromodynamics (QCD), the color degree of freedom of quarks is described by the fundamental representation $\bm{3}$ of the SU(3) gauge group. The tensor product of two color triplets decomposes into the direct sum $\textbf{3}\otimes\bm{3}=\bar{\bm{3}}\oplus\bm{6}$, where the antisymmetric part yields the color antitriplet $\bar{\bm{3}}$, and the symmetric part corresponds to the color sextet $\bm{6}$. When combined with the two possible S-wave spin configurations, the antisymmetric spin-singlet $^1S_0$ and the symmetric spin-triplet $^3S_1$, leads to four distinct diquark configurations: [$^3S_1\mbox{-}\bar{\bm{3}}$], [$^1S_0\mbox{-}\bm{6}$], [$^3S_1\mbox{-}\bm{6}$], [$^1S_0\mbox{-}\bar{\bm{3}}$]. For a diquark composed of two quarks of the same flavor (e.g., $cc$ or $bb$), the fermion spin-statistic principle enforces overall antisymmetry under exchange of the two identical fermions. Consequently, only the configurations that are antisymmetric in the combined spin–color space are allowed, namely [$^3S_1\mbox{-}\bar{\bm{3}}$] and [$^1S_0\mbox{-}\bm{6}$]. In contrast, for a diquark consisting of quarks with different flavors (e.g., $bc$), no such symmetry constraint applies, and all four spin–color configurations are physically allowed. 

\begin{figure}[htbp!]			
	\centering
	\caption{Two representative leading-order Feynman diagrams for the radiative decay $T_{4c} \to J/\psi+\gamma$ of the fully-charmed tetraquark, corresponding to the gluon-exchange (left) and quark-rearrangement (right) decay topologies respectively.}
	\subfigure{\includegraphics[scale=0.4]{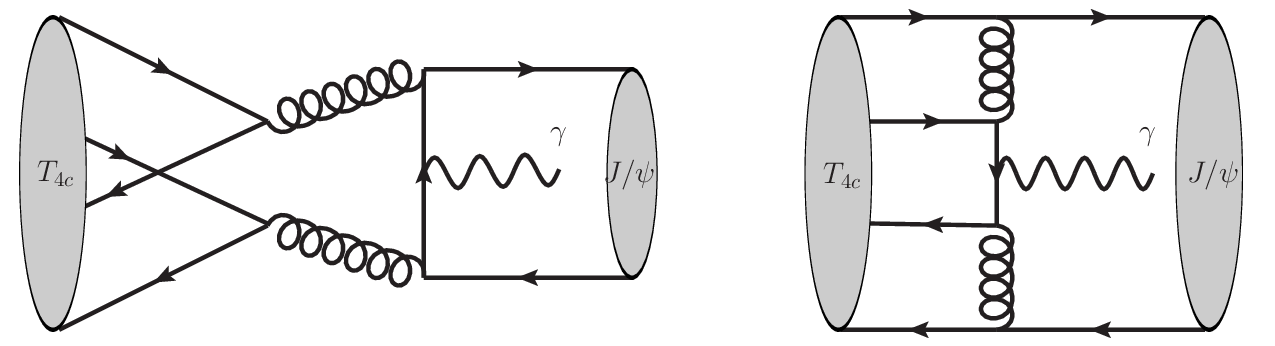}}
	\label{Fig_T4c2Jpsir}
\end{figure}
For simplicity, we will illustrate our theoretical framework using the $T_{4c} \to J/\psi\gamma$ process as an example, other processes can be obtained through an analogous way. In Fig. \ref{Fig_T4c2Jpsir}, we present the two representative leading-order Feynman diagrams for the radiative decay $T_{4c} \to J/\psi+\gamma$. The left diagram depicts the gluon exchange topology: two gluons mediate the strong interaction between $T_{4c}$ and $J/\psi$, while the photon is radiated from the intermediate quark line. The right diagram corresponds to the quark rearrangement topology: the four quarks ($cc\bar{c}\bar{c}$) of the initial $T_{4c}$ rearrange their configuration, with a photon emitted from the intermediate quark line during the process. In the decay of $T_{4c/b}$, there are a total of 220 Feynman diagrams, including 28 for gluon-exchange and 192 for quark-rearrangement processes, while for the $T_{bc\bar{b}\bar{c}}$ decay, 62 diagrams are involved in total.

Within the NRQCD factorization framework, the radiative decay width of fully charm tetraquark into $J/\psi+\gamma$ can be formulated as
\begin{align}
	\Gamma[T_{4c}\to J/\psi\gamma] = \sum_{n} \hat{\Gamma}[T_{4c}\to J/\psi\gamma] \braket{\mathcal{O}^{T}_{n}}\braket{\mathcal{O}^{J/\psi}},
\end{align}
where $\braket{\mathcal{O}^{T}_{n}}$ are the long-distance matrix elements (LDMEs)
describing the possibility of finding the intermediate state with color-spin configuration $n$ inside the tetraquark T, $\braket{\mathcal{O}^{J/\psi}}$ is the LDME for $J/\psi$, and
$\hat{\Gamma}[T_{4c}\to J/\psi\gamma]$ is the short distance coefficient (SDC), a quantity that can be evaluated perturbatively. For fully charm tetraquark, the possible configurations $n$ are
\begin{align*}
	&0^{++}[(\bm{6}\mbox{-}{^1S_0})\otimes(\bar{\bm{6}}\mbox{-}^1S_0)], \hspace{0.5cm} 0^{++}[(\bar{\bm{3}}\mbox{-}{^3S_1})\otimes(\bm{3}\mbox{-}^3S_1)],\\
	&1^{+-}[(\bar{\bm{3}}\mbox{-}{^3S_1})\otimes(\bm{3}\mbox{-}^3S_1)], \hspace{0.5cm} 2^{++}[(\bar{\bm{3}}\mbox{-}{^3S_1})\otimes(\bm{3}\mbox{-}^3S_1)].
\end{align*}
 
The corresponding operators are defined as:
\begin{align}
	&\mathcal{O}^{(0)}_{\bm{6}\otimes\bar{\bm{6}}} = [\psi_{a}^T(i\sigma^2)\psi_b][\chi_c^{\dagger}(i\sigma^2)\chi_d^*]\mathcal{C}_{\bm{6}\otimes\bar{\bm{6}}}^{ab;cd},\nonumber\\
	&\mathcal{O}^{(0)}_{\bar{\bm{3}}\otimes\bm{3}} = -\dfrac{1}{\sqrt{3}} [\psi_{a}^T(i\sigma^2)\sigma^i\psi_b][\chi_c^{\dagger}\sigma^i(i\sigma^2)\chi_d^*]\mathcal{C}_{\bar{\bm{3}}\otimes\bm{3}}^{ab;cd},\nonumber\\
	&\mathcal{O}^{(1)}_{\bar{\bm{3}}\otimes\bm{3}} = -\dfrac{i}{\sqrt{2}}\epsilon_{H}^{i*} [\psi_{a}^T(i\sigma^2)\sigma^j\psi_b][\chi_c^{\dagger}\sigma^k(i\sigma^2)\chi_d^*]\epsilon^{ijk}\mathcal{C}_{\bar{\bm{3}}\otimes\bm{3}}^{ab;cd},\nonumber\\
	&\mathcal{O}^{(2)}_{\bar{\bm{3}}\otimes\bm{3}} = \dfrac{1}{2}\epsilon_{H}^{ij*} [\psi_{a}^T(i\sigma^2)\sigma^k\psi_b][\chi_c^{\dagger}\sigma^\ell(i\sigma^2)\chi_d^*]\Gamma^{ij;k\ell}\mathcal{C}_{\bar{\bm{3}}\otimes\bm{3}}^{ab;cd},
\end{align}
with the LDMEs defined as $\braket{\mathcal{O}^{(J)}_{\bar{\bm{3}}\otimes\bm{3}}} = \big|\bra{0}\mathcal{O}^{(J)}_{\bar{\bm{3}}\otimes\bm{3}}\ket{T_{4c}}\big|^2$ and $\braket{\mathcal{O}^{(0)}_{\bm{6}\otimes\bar{\bm{6}}}} = \big|\bra{0}\mathcal{O}^{(0)}_{\bm{6}\otimes\bar{\bm{6}}}\ket{T_{4c}}\big|^2$. In this expression, $\psi$ and $\chi^\dagger$ correspond to Pauli spinor fields that annihilate heavy quark and heavy antiquark respectively, while $\sigma^i$ denotes the standard 2-dimensional Pauli matrix. The quantities $\epsilon^{i/ij}_H$ is the polarization vector (tensor) of the spin-1 (spin-2) tetraquark state. We explicitly define the color projection operators $\mathcal{C}$ and the symmetric traceless polarization tensor $\Gamma$ as follows:
\begin{align}
	\mathcal{C}^{ab;cd}_{\bar{\bm{3}}\otimes \bm{3}} = \dfrac{\delta^{ac}\delta^{bd} - \delta^{ad}\delta^{bc}}{2\sqrt{3}},\hspace{0.2cm} \mathcal{C}^{ab;cd}_{\bm{6}\otimes \bar{\bm{6}}} = \dfrac{\delta^{ac}\delta^{bd} + \delta^{ad}\delta^{bc}}{2\sqrt{6}},\hspace{0.2cm} \Gamma^{ij;k\ell} = \delta^{ik}\delta^{j\ell} + \delta^{i\ell}\delta^{jk} - \frac{2}{3}\delta^{ij}\delta^{k\ell},
\end{align}
where $a\mbox{-}d$ are the SU(3) color indices of the constituent diquarks and the antidiquarks respectively, and $i\mbox{-}\ell\ (=1,2,3)$ denote Cartesian spatial indices.

To project the product of two fermion bilinears into a diquark state with definite spin and color quantum numbers, we perform a standard transformation of the Dirac spinor chain to rewrite the product as a compact trace over Dirac indices, which greatly simplifies subsequent amplitude calculations. The manipulation proceeds as:
\begin{align}
	  \bar{v}(p_c)\Gamma u(p_{\bar{c}}) \bar{v}(p_{c'})\Gamma' u(p_{\bar{c'}}) = & \bar{v}(p_c)\Gamma u(p_{\bar{c}}) u^T(p_{\bar{c'}})\Gamma'^T\bar{v}^T(p_{c'})\nonumber\\
	= & \bar{v}(p_c)\Gamma u(p_{\bar{c}}) u^T(p_{\bar{c'}})\mathcal{C}\mathcal{C}^{-1}\Gamma'^T\mathcal{C}\mathcal{C}^{-1}\bar{v}^T(p_{c'})\nonumber\\
	= & -\bar{v}(p_c)\Gamma u(p_{\bar{c}}) \bar{v}(p_{\bar{c'}})\Gamma'^{\mathcal{C}}u(p_{c'})\nonumber\\
	\to & -{\rm Tr}[\Pi^{0/1}_{cc'}\Gamma\Pi^{0/1}_{\bar{c}\bar{c}'}\Gamma'^{\mathcal{C}}]\mathcal{C}^{ab;cd}.
\end{align}
This manipulation relies on the standard properties of the Dirac charge conjugation matrix, defined in the conventional Dirac gamma basis as $\mathcal{C} = -i\gamma^2\gamma^0$. For an arbitrary matrix $\Gamma$, the charge-conjugated is defined as $\Gamma^{\mathcal{C}} \equiv \mathcal{C}^{-1}\Gamma^T\mathcal{C}$, and the charge conjugation matrix satisfies the following core identities:
\begin{align}
	\mathcal{C}\mathcal{C}^{-1} = 1, \mathcal{C}^{-1}\gamma^{\mu T}\mathcal{C} = -\gamma^{\mu}, u^T(p,\sigma)\mathcal{C} = -\bar{v}(p,\sigma), \mathcal{C}^{-1}\bar{v}^T(p,\sigma) = u(p,\sigma).
\end{align}

The corresponding spin projectors \footnote{We project $u(p_c')\bar{v}(p_c)$ into spin singlet (s=0) or spin triplet (s=1), for $u(p_c)\bar{v}(p_c')$ projection, an extra $(-1)^{s+1}$ factor is needed, see \cite{Zhang:2023ffe} for details. The projectors for the production of $J/\psi$ and $\eta_c$ are similar to those after complex conjugation and including the color factor $\delta^{ab}/\sqrt{N_c}$.}
\begin{align}
	\Pi^{0} = \dfrac{1}{2\sqrt{2m_c}}\gamma^5(\frac{\slashed{P}}{2}-2m_c),\hspace{1cm} \Pi^{1} = \dfrac{1}{2\sqrt{2m_c}}\slashed{\epsilon}(\frac{\slashed{P}}{2}-2m_c).
\end{align}
To generalize these projection operators for a $bb\ (bc)$-diquark, one simply makes the replacement $2m_c \to 2m_b\ (m_b+m_c)$ in the expressions above.

To construct a tetraquark state from two spin-triplet diquark and anti-diquark in a fully covariant framework, we first couple the two spin-1 diquark and anti-diquark into states with total spin $J = 0, 1$ and $2$. This spin decomposition is systematically implemented via a set of orthogonal, properly normalized covariant spin projectors $J_{0,1,2}^{\mu\nu}$, whose explicit forms are
\begin{align}
	J^{\mu\nu}_{0} = \dfrac{1}{\sqrt{3}}\eta^{\mu\nu},\hspace{1.1cm} J^{\mu\nu}_{1} = -\dfrac{i}{\sqrt{2P^2}}\epsilon^{\mu\nu\alpha\beta}\epsilon_{\alpha}P_\beta,\nonumber \\
	J^{\mu\nu}_{2} = \epsilon_{\alpha\beta}\left\{ \dfrac{1}{2}\left[ \eta^{\mu\alpha}\eta^{\nu\beta}+\eta^{\mu\beta}\eta^{\nu\alpha}\right] - \dfrac{1}{3}\eta^{\mu\nu}\eta^{\alpha\beta} \right\},		
\end{align}
where $P^{\mu}$ is the total four-momentum of the tetraquark system. The tensor $\eta^{\mu\nu} = -g^{\mu\nu} + \frac{P^{\mu}P^{\nu}}{P^2}$ defines the transverse projector relative to $P^{\mu}$. By construction, $\epsilon_{\alpha}$ is the polarization vector for the resulting total spin-1 tetraquark, while $\epsilon_{\alpha\beta}$ is the symmetric polarization tensor for the total spin-2 tetraquark, satisfy the standard polarization sum:
\begin{align}
	&\sum_{\lambda} \epsilon^{\alpha}(P,\lambda)\epsilon^{\alpha'}(P,\lambda) = \eta^{\alpha\alpha'},\nonumber\\
	&\sum_{\lambda} \epsilon^{\alpha\beta}(P,\lambda)\epsilon^{\alpha'\beta'}(P,\lambda) = \dfrac{1}{2}\left[ \eta^{\alpha\alpha'}\eta^{\beta\beta'}+\eta^{\alpha\beta'}\eta^{\beta\alpha'}\right] - \dfrac{1}{3}\eta^{\alpha\beta}\eta^{\alpha'\beta'}.
\end{align}

Finally, the SDC for fully charm tetraquark is formulated as
\begin{align}
	&\hat{\Gamma}[T^{(J)}_{4c}\to J/\psi(\eta_c)\gamma] = \dfrac{3}{256m_c(2J+1)\pi} \Big| \mathcal{M}(cc\bar{c}\bar{c}[\mathcal{H}_{cc}\otimes\mathcal{H}_{\bar{c}\bar{c}}] \to c\bar{c}[\bm{1}]+\gamma) \Big|^2,\nonumber \\
	&\hat{\Gamma}[T^{(J)}_{4b}\to \Upsilon(\eta_b)\gamma] = \dfrac{3}{256m_b(2J+1)\pi} \Big| \mathcal{M}(bb\bar{b}\bar{b}[\mathcal{H}_{bb}\otimes\mathcal{H}_{\bar{b}\bar{b}}] \to b\bar{b}[\bm{1}]+\gamma) \Big|^2,\nonumber \\
	&\hat{\Gamma}[T^{(J)}_{bc\bar{b}\bar{c}}\to J/\psi(\eta_c)\gamma] = \dfrac{m_b(m_b+2m_c)}{32(m_b+m_c)^3(2J+1)\pi} \Big| \mathcal{M}(bc\bar{b}\bar{c}[\mathcal{H}_{bc}\otimes\mathcal{H}_{\bar{b}\bar{c}}] \to c\bar{c}[\bm{1}]+\gamma) \Big|^2,\nonumber \\
	&\hat{\Gamma}[T^{(J)}_{bc\bar{b}\bar{c}}\to \Upsilon(\eta_b)\gamma] = \dfrac{m_c(m_c+2m_b)}{32(m_b+m_c)^3(2J+1)\pi} \Big| \mathcal{M}(bc\bar{b}\bar{c}[\mathcal{H}_{bc}\otimes\mathcal{H}_{\bar{b}\bar{c}}] \to b\bar{b}[\bm{1}]+\gamma) \Big|^2,  
\end{align}
where $\mathcal{H}_{QQ',\bar{Q}\bar{Q'}}$ denote the color-spin configurations for heavy diquark. In the calculations, we generate the Feynman diagrams and corresponding amplitudes using FeynArts \cite{Hahn:2000kx}, and perform subsequent algebraic manipulations with FeynCalc \cite{Mertig:1990an}. To verify the correctness of our results, we reproduced the analytical results for similar processes in Refs. \cite{Zhang:2023ffe,Sang:2023ncm,Feng:2023agq} using our code.

\section{NUMERICAL RESULTS}
In this section, we present numerical results for the width of fully heavy tetraquark decay into photon plus quarkonium, including $T_{4c/4b}^{0/2++} \to \gamma J/\psi/\Upsilon$, $T_{4c/4b}^{1+-} \to \gamma\eta_c/\eta_b$, $T_{bc\bar{b}\bar{c}}^{0/1/2++} \to \gamma J/\psi(\Upsilon)$ and $T_{bc\bar{b}\bar{c}}^{1+-} \to \gamma \eta_c(\eta_b)$ within the following parameters:
\begin{align}
	\alpha = 1/137.065,\hspace{1.12cm} m_c = 1.5\ {\rm GeV},\hspace{1.12cm} m_b = 4.9\ {\rm GeV},\nonumber\\
	\alpha_s(2m_c) = 0.249,\ \alpha_s(m_b+m_c) = 0.199,\ \alpha_s(2m_b) = 0.179,\nonumber\\
	\braket{\mathcal{O}^{J/\psi}[^3S_1]} = 1.32\ \rm GeV,\hspace{1.9cm} \braket{\mathcal{O}^{\Upsilon}[^3S_1]} = 9.28\ \rm GeV.
\end{align} 
Here the value of the strong coupling constant is obtained from the software RunDec. \cite{Chetyrkin:2000yt} with three typical conventions $\mu = 2m_c$, $m_b+m_c$ and $2m_b$ for $T_{4c}$, $T_{bc\bar{b}\bar{c}}$ and $T_{4b}$ states, respectively. The LDMEs for the production of $J/\psi$ and $\Upsilon$ are taken from Refs.\cite{Bodwin:2007fz,Li:2019anc}, the LDMEs for $\eta_c$ and $\eta_b$ are 1/3 of $J/\psi$ and $\Upsilon$ according to heavy quark spin symmetry. The LDMEs for $T_{4c}$ are taken from two phenomenological potential models based on the extended relativized quark model (ERQM, Model I) \cite{Lu:2020cns} and the nonrelativistic quark model (NRQM, Model II) \cite{liu:2020eha}, which rely on Cornell type potential in four-body Schr\"odinger equation, the LDMEs are listed in Table \ref{Tab_LDME_T4c}. 
\begin{table}[htb]
	\centering
	\caption{Numerical values of the LDMEs for $T_{4c}$ in Model I-II, in units of $\rm Gev^{9}$.}
	\setlength{\tabcolsep}{12pt}
	\begin{tabular}{|c!{\vrule}c|c|c|c|c|}
		\hline
		$\rm J^{PC}$ & \multicolumn{3}{c|}{$0^{++}$} & $1^{+-}$ & $2^{++}$ \\
		\hline
		LDMEs & $\braket{ \mathcal{O}^{(0)}_{\bar{\bm{3}} \otimes \bm{3}} }$ & $\braket{ \mathcal{O}^{(0)}_{\rm mix} }$ & $\braket{ \mathcal{O}^{(0)}_{\bm{6} \otimes \bar{\bm{6}}} }$ & $\braket{ \mathcal{O}^{(1)}_{\bar{\bm{3}} \otimes \bm{3}} }$ & $\braket{ \mathcal{O}^{(2)}_{\bar{\bm{3}} \otimes \bm{3}} }$ \\
		\hline
		Model I & $0.0347$ & $0.0211$ & $0.0128$ & $0.0260$ & $0.0144$ \\
		\hline
		Model II & $0.0187$ & $-0.0161$ & $0.0139$ & $0.0160$ & $0.0126$ \\
		\hline
	\end{tabular}
    \label{Tab_LDME_T4c}
\end{table}

To estimate the LDMEs for $T_{4b}$ and $T_{bc\bar{b}\bar{c}}$, which are absent in current literature,
the NRQCD long-distance matrix elements governing tetraquark transitions can be represented in terms of the tetraquark wavefunctions at the origin \cite{Feng:2023agq},
\begin{align}
	\braket{ \mathcal{O}^{(J)}_{\bm{C_1}\otimes \bm{C_2}} } \approx 16(2J+1)\psi_{\bm{C_1}}(0)\psi_{\bm{C_2}}^*(0), 
\end{align}
where $\bm{C_{1/2}}$ is the color configuration ($\bar{\bm{3}}$ or $\bm{6}$) for the diquark. The exact value for $\psi_C$ represents a complex four-body problem in quantum chromodynamics. To render this computationally tractable, we employ a simplified approach by decomposing the system, as employed in Refs.\cite{Debastiani:2017msn,Liang:2025wbt}, into three effective two-body motions: I. the internal dynamics of the diquark; II. the internal dynamics of the antidiquark; III. the relative motion between these two composite color sources. Under this approximation, the wave function at the origin $\psi_C(0)$ can be factorized as the product of individual components: $\psi_{QQ'}\psi_{\bar{Q}\bar{Q}'}\psi_{D\bar{D}}$, where $\psi_{D\bar{D}}$ describes the diquark-antidiquark relative motion. And the LDMEs can be related to radical wave functions at the origin
\begin{align}
	\braket{ \mathcal{O}^{(J)}_{\bm{C_1} \otimes \bm{C_2}} } = \dfrac{2J+1}{4\pi^3} |R^{\bm{C_1}}_{QQ'}(0)|^2 |R^{\bm{C_2}}_{\bar{Q}\bar{Q}'}(0)|^2 |R^{\bm{C_1} \otimes \bm{C_2}}_{D\bar{D}}(0)|^2,
\end{align}
where the wave functions evaluated by solving the Schr\"odinger equation with a Cornell-inspired potential \cite{Debastiani:2017msn,Liang:2025wbt}. 

In our estimation, we use the nonrelativistic potential contains the one-gluon exchange (OGE) with linear confining term and the smeared spin-spin (SS) interaction:
\begin{align}
	V^{\rm OGE}(r) = \kappa_s \dfrac{\alpha_s}{r} + br,\hspace{1cm} V^{\rm SS}(r) = -\dfrac{8\pi\kappa_s\alpha_s}{3m_1m_2}(\dfrac{\sigma}{\sqrt{\pi}})^3\exp^{-\sigma^2r^2}\bm{S_1}\cdot\bm{S_2}
\end{align}
In this expression, $\kappa_s$, often referred to as the color factor, is a dimensionless parameter fully determined by the specific color representation configuration of the studied system, and it can take either negative or positive values, corresponding to net attractive or repulsive color interaction respectively. $\alpha_s$ is the QCD strong coupling, $b$ is the parameter that quantifies the strength of the color confining interaction between the two colored constituents, and $\sigma$ is the parameter for the Gaussian function.

\begin{table}[htb]
	\centering
	\caption{The squared radial wave function at origin $|R(0)|^2$ ($\rm GeV^3$) for fully heavy tetraquark.}
	\begin{tabular}{|m{1.5cm}<{\centering}|m{2.6cm}<{\centering}|m{2.6cm}<{\centering}|m{2.6cm}<{\centering}|m{2.6cm}<{\centering}|}
		\hline
		$|R(0)|^2$  & $QQ'[\bar{\bm{3}}]$ & $QQ'[\bm{6}]$ & $QQ'[\bar{\bm{3}}]\otimes\bar{Q}\bar{Q'}[\bm{3}]$ & $QQ'[\bm{6}]\otimes\bar{Q}\bar{Q'}[\bar{\bm{6}}]$ \\
		\hline
		$cc\bar{c}\bar{c}$ & 0.315 & 0.272 & 8.95 & 71.13 \\
		\hline
		$bc\bar{b}\bar{c}$ & 0.549 & 0.328 & 22.69 & 284.4 \\
		\hline
		$bb\bar{b}\bar{b}$ & 1.226 & 0.748 & 26.38 & 327.3 \\
		\hline
	\end{tabular}
	\label{Tab_R0}
\end{table}

\begin{table}[htb]
	\centering
	\caption{Numerical values of the LDMEs ($\rm Gev^{9}$) for $T_{4b}$.}
	\setlength{\tabcolsep}{12pt}
	\begin{tabular}{|c!{\vrule}c|c|c|c|c|}
		\hline
		$\rm J^{PC}$ & \multicolumn{3}{c|}{$0^{++}$} & $1^{+-}$ & $2^{++}$ \\
		\hline
		LDMEs & $\braket{ \mathcal{O}^{(0)}_{\bar{\bm{3}} \otimes \bm{3}} }$ & $\braket{ \mathcal{O}^{(0)}_{\rm mix} }$ & $\braket{ \mathcal{O}^{(0)}_{\bm{6} \otimes \bar{\bm{6}}} }$ & $\braket{ \mathcal{O}^{(1)}_{\bar{\bm{3}} \otimes \bm{3}} }$ & $\braket{ \mathcal{O}^{(2)}_{\bar{\bm{3}} \otimes \bm{3}} }$ \\
		\hline
		$\braket{\mathcal{O}_{T_{4c}}}$ & 0.00716 & 0.0174 & 0.0424 & 0.0215 & 0.0358 \\
		\hline
		$\braket{\mathcal{O}_{T_{4b}}}$ & 0.319 & 0.686 & 1.476 & 0.959 & 1.599 \\
		\hline
	\end{tabular}
	\label{Tab_LDME_T4b}
\end{table}

Since the interaction in the color sextet $\bm{6}$ configuration is repulsive and cannot support a stable bound state, we estimate $R^{\bm{6}}_{QQ'}(0)$ following the NRQCD scaling rule given in Ref.\cite{Ma:2003zk}: $|R^{\bm{6}}_{QQ'}(0)|^2 = v^2 |R_{Q\bar{Q'}}(0)|^2$
where we take $v^2 = 0.3$, $0.2$ and $0.1$ for the $c\bar{c}$, $b\bar{c}$ and $b\bar{b}$ systems, respectively. For our input parameters, we adopt $|R_{c\bar{c}}(0)|^2 = 0.907\ \rm GeV^3$, $|R_{b\bar{c}}(0)|^2 = 1.642\ \rm GeV^3$ and $|R_{b\bar{b}}(0)|^2 = 7.48\ \rm GeV^3$. The values of the radial wave function at the origin for heavy diquark and diquark-diquark systems are summarized in Table \ref{Tab_R0} where the heavy quark spin symmetry is applied. 

\begin{table}[htb]
	\centering
	\caption{The estimated LDMEs (in unit $\rm GeV^9$) for $(0,1,2)^{++}$ $T_{bc\bar{b}\bar{c}}$.}
	\begin{tabular}{|m{1.3cm}<{\centering}|m{1.6cm}<{\centering}|m{1.6cm}<{\centering}|m{1.6cm}<{\centering}|m{1.6cm}<{\centering}|m{1.8cm}<{\centering}|m{1.8cm}<{\centering}|m{1.6cm}<{\centering}|m{1.6cm}<{\centering}|}
		\hline
		\multirow{2}{*}{$\rm J^{PC}$} & \multicolumn{4}{c|}{$0^{++}$} & \multicolumn{2}{c|}{$1^{++}$} & \multicolumn{2}{c|}{$2^{++}$} \\
		\cline{2-9}
		& {\footnotesize $(bc)_{\bm{6}}^{0}\mbox{-}(\bar{b}\bar{c})_{\bar{\bm{6}}}^{0}$ } & {\footnotesize $(bc)_{\bm{6}}^{1}\mbox{-}(\bar{b}\bar{c})_{\bar{\bm{6}}}^{1}$ } & {\footnotesize $(bc)_{\bar{\bm{3}}}^{0}\mbox{-}(\bar{b}\bar{c})_{\bm{3}}^{0}$ } & {\footnotesize $(bc)_{\bar{\bm{3}}}^{1}\mbox{-}(\bar{b}\bar{c})_{\bm{3}}^{1}$ } & {\footnotesize $(bc)_{\bm{6}}^{1}\mbox{-}(\bar{b}\bar{c})_{\bar{\bm{6}}}^{0A}$ } & {\footnotesize $(bc)_{\bar{\bm{3}}}^{1}\mbox{-}(\bar{b}\bar{c})_{\bm{3}}^{0A}$ } & {\footnotesize $(bc)_{\bm{6}}^{1}\mbox{-}(\bar{b}\bar{c})_{\bar{\bm{6}}}^{1}$ } & {\footnotesize $(bc)_{\bar{\bm{3}}}^{1}\mbox{-}(\bar{b}\bar{c})_{\bm{3}}^{1}$ } \\
		\hline
		$\braket{\mathcal{O}_{T_{bc\bar{b}\bar{c}}}}$ & 0.246 & 0.246 & 0.0551 & 0.0551 & 0.738 & 0.165 & 1.23 & 0.275\\
		\hline
	\end{tabular}
	\label{Tab_LDME_Tbcbc++}
\end{table}

\begin{table}[htb]
	\centering
	\caption{The estimated LDMEs (in unit $\rm GeV^9$) for $1^{+-}$ $T_{bc\bar{b}\bar{c}}$.}
	\begin{tabular}{|m{1.3cm}<{\centering}|m{1.6cm}<{\centering}|m{1.6cm}<{\centering}|m{1.8cm}<{\centering}|m{1.8cm}<{\centering}|}
		\hline
		\multirow{2}{*}{$\rm J^{PC}$} & \multicolumn{4}{c|}{$1^{+-}$} \\
		\cline{2-5}
		& {\footnotesize $(bc)_{\bm{6}}^{1}\mbox{-}(\bar{b}\bar{c})_{\bar{\bm{6}}}^{1}$ } &  {\footnotesize $(bc)_{\bar{\bm{3}}}^{1}\mbox{-}(\bar{b}\bar{c})_{\bm{3}}^{1}$ } &  {\footnotesize $(bc)_{\bm{6}}^{1}\mbox{-}(\bar{b}\bar{c})_{\bar{\bm{6}}}^{0S}$ } & {\footnotesize $(bc)_{\bar{\bm{3}}}^{1}\mbox{-}(\bar{b}\bar{c})_{\bm{3}}^{0S}$ } \\
		\hline
		$\braket{\mathcal{O}_{T_{bc\bar{b}\bar{c}}}}$ & 0.738 & 0.165 & 0.738 & 0.165\\
		\hline
	\end{tabular}
	\label{Tab_LDME_Tbcbc+-}
\end{table}
Finally, the resulting LDMEs for the tetraquark are listed in Table \ref{Tab_LDME_T4b}-\ref{Tab_LDME_Tbcbc+-}. As can be seen, the roughly estimated LDMEs for the $T_{4c}$ state are several times larger or smaller than the corresponding results from Models I and II.

\subsection{$T_{4c}$ decays into charmonium plus photon}
We present the calculation of the squared amplitudes for the fully charm tetraquark decay $T_{4c} \to \gamma J/\psi (\eta_c)$ at leading order. This is performed within the NRQCD factorization framework using a double expansion in the strong coupling constant and the heavy quark velocity. The analytic results are given in Appendix A. The topologies in this channel are categorized into two distinct structures: the gluon-exchange and quark-rearrangement. In the first mode, both $c\bar{c}$ pairs must annihilate at a short distance, subsequently producing a new $c\bar{c}$ pair. Consequently, its probability is significantly suppressed compared to the quark-rearrangement mode, where only one pair annihilates. Our calculation confirms this intuitive picture. We find that the contribution from gluon-exchange diagrams is negligible, accounting for less than 0.4\% of the total width. Extending this analysis, we identify a clear decay hierarchy based on topological suppression: double-rearrangement, where two initial $c\bar{c}$ pairs directly combine into two charmonia, charmonium+$D\bar{D}$ or 4 $D$-mesons; single-rearrangement, where one $c\bar{c}$ pair forms a charmonium or two $D$-meson; gluon exchange, involving the annihilation of both $c\bar{c}$ pairs into gluons, which then hadronize into light hadrons; and photon exchange, where annihilation yields direct photon pairs. We list the typical decay channels and the estimated widths in Table \ref{Tab_Width_T4c2All} to confirm our assertion.
\begin{table}[ht]
	\caption{Various decay widths of $T_{4c}$ in Refs: di-charmonium decay\cite{liu:2020eha,Lu:2025lyu,Xia:2025mgk}, hadronic decay \cite{Zhang:2023ffe} (We take Model-I LDMEs for numerical results), di-photon decay \cite{Biloshytskyi:2022dmo,Biloshytskyi:2022pdl,Liu:2025mxv}.}
	\begin{center}
		\centering
		\begin{tabular}{|m{1.5cm}<{\centering}|m{1.5cm}<{\centering}|m{1.5cm}<{\centering}|m{2.cm}<{\centering}|m{1.6cm}<{\centering}|m{1.6cm}<{\centering}|m{1.6cm}<{\centering}|m{2.5cm}<{\centering}|}
			\toprule
			\hline
			$\Gamma(\ \rm MeV)$ & $\eta_c\eta_c$ & $\eta_c J/\psi$ & $J/\psi J/\psi$ & $c\bar{c}$ & $gg$& $qq$ & $\gamma\gamma$  \\
			\hline
			$0^{++}_{\rm I}$  & \makecell{1.45 \cite{liu:2020eha}\\ 4.53 \cite{Xia:2025mgk}} &  - &  0.7 \cite{liu:2020eha} & 0.001 \cite{Zhang:2023ffe} & 0.033 \cite{Zhang:2023ffe} & - & $6\times10^{-4}$ \cite{Liu:2025mxv}\\
			\hline
			$0^{++}_{\rm II}$ & \makecell{0.12 \cite{liu:2020eha}\\ 6.95 \cite{Xia:2025mgk}} & -  & \makecell{1.78 \cite{liu:2020eha}\\ 4.14 \cite{Xia:2025mgk}} & 0.208 \cite{Zhang:2023ffe} & 0.363 \cite{Zhang:2023ffe} & - & $6\times10^{-4}$ \cite{Liu:2025mxv}\\
			\hline
			$1^{+-}$ & - & \makecell{0.45 \cite{liu:2020eha}\\ 4.31 \cite{Xia:2025mgk}} & - & - & - & - & - \\
			\hline
			$2^{++}$ & - & - & \makecell{0.36 \cite{liu:2020eha}\\ 10-80 \cite{Lu:2025lyu}\\ 3.14 \cite{Xia:2025mgk}} & 1.57 \cite{Zhang:2023ffe} & 0.097 \cite{Zhang:2023ffe} & 0.286 \cite{Zhang:2023ffe} & \makecell{$6.7\times10^{-2}$ \cite{Biloshytskyi:2022dmo}\\ $1\times10^{-2}$ \cite{Biloshytskyi:2022pdl}\\ $1\times10^{-4}$ \cite{Liu:2025mxv}}\\
			\hline		
		\end{tabular}
	\end{center}    
	\label{Tab_Width_T4c2All}
\end{table}

It should be noted that two configurations, $(cc)^{1}_{\bar{\bm{3}}}\otimes(\bar{c}\bar{c})^{1}_{\bm{3}}$ and $(cc)^{0}_{\bm{6}}\otimes(\bar{c}\bar{c})^{0}_{\bar{\bm{6}}}$, share the exact same quantum numbers for the $T_{4c}^{0++}$ state. Consequently, the physical states must be two orthogonal mixtures of these configurations, which can be parameterized as
\begin{align}
	\begin{pmatrix}
		\ket{0^{++}_{\rm I}} \\
		\ket{0^{++}_{\rm II}}
	\end{pmatrix} =
	\begin{pmatrix}
		\cos\theta_c & \sin\theta_c \\
		\sin\theta_c & -\cos\theta_c
	\end{pmatrix} 
	\begin{pmatrix}
		\ket{0^{++}_{(cc)^{1}_{\bar{\bm{3}}}\otimes(\bar{c}\bar{c})^{1}_{\bm{3}}}} \\
		\ket{0^{++}_{(cc)^{0}_{\bm{6}}\otimes(\bar{c}\bar{c})^{0}_{\bar{\bm{6}}}}}
	\end{pmatrix},
\end{align}
where the mixture angle $\theta_c \approx 52^\circ$ ($\theta_b \approx 66^\circ$ for $T_{4b}^{0++}$ states) \cite{Lu:2020cns,liu:2020eha}. Hence the measured decay width should be written as
\begin{align}
	\Gamma[T_{4c}^{0^{++}_{\rm I/II}} \to J/\psi\gamma] &= \dfrac{3\braket{\mathcal{O}^{J/\psi}}}{256m_c\pi}\Big( \cos^2\theta_c \big| \mathcal{M}^{0++}_{\bar{\bm{3}}\otimes\bm{3}} \big|^2 \braket{\mathcal{O}_{\bar{\bm{3}}\otimes\bm{3}}^{0++}} + \sin^2\theta_c \big| \mathcal{M}^{0++}_{\bm{6}\otimes\bar{\bm{6}}} \big|^2 \braket{\mathcal{O}_{\bm{6}\otimes\bar{\bm{6}}}^{0++}} \nonumber\\
	&\pm 2\sin\theta_c\cos\theta_c \big| \mathcal{M}^{0++}_{\rm mix} \big|^2 \braket{\mathcal{O}_{\rm mix}^{0++}} \Big).
\end{align}
While the tensor ($2^{++}$) and axial-vector ($1^{+-}$) states exhibit simpler decay patterns:
\begin{align}
	\Gamma[T_{4c}^{2++}\to J/\psi\gamma] &= \dfrac{3\braket{\mathcal{O}^{J/\psi}}}{1280m_c\pi} |\mathcal{M}_{\bar{\bm{3}}\otimes\bm{3}}^{2++}|^2\braket{\mathcal{O}^{2++}_{\bar{\bm{3}}\otimes\bm{3}}}, \\
	\Gamma[T_{4c}^{1+-}\to \eta_c\gamma] &= \dfrac{\braket{\mathcal{O}^{\eta_c}}}{256m_c\pi} |\mathcal{M}_{\bar{\bm{3}}\otimes\bm{3}}^{1+-}|^2\braket{\mathcal{O}^{1+-}_{\bar{\bm{3}}\otimes\bm{3}}}.
\end{align}

\begin{table}[ht]
	\caption{The decay widths (in unit MeV) of $T_{4c} \to \gamma+J/\psi(\eta_c)$ with the Model-I and Model-II LDMEs (in brackets).}
	\begin{center}
		\centering
		\begin{tabular}{|m{2.cm}<{\centering}|m{2.2cm}<{\centering}|m{2.2cm}<{\centering}|m{2.2cm}<{\centering}|m{2.2cm}<{\centering}|}
			\toprule
			\hline
			$\Gamma(\ \rm MeV)$ & $0^{++}_{\rm I}$ & $0^{++}_{\rm II}$ & $1^{+-}$ & $2^{++}$ \\
			\hline			
			$\gamma J/\psi$     & 1.06 (0.885)           & 1.53 (0.523)             & -        & 0.132 (0.116)    \\
			\hline
			$\gamma\eta_c$      & -                & -                 &  1.02 (0.628)    & -        \\
			\hline  		
		\end{tabular}
	\end{center}    
	\label{Tab_Width_T4c2Jpsir}
\end{table}
After considering the interference, and takes the LDMEs provided above, we estimate the decay widths in Table \ref{Tab_Width_T4c2Jpsir}. The decay widths for the two $0^{++}$ tetraquark are around 1 MeV. The mass of $0^{++}$ ground state varies from 5966 MeV \cite{Berezhnoy:2011xn} to 7016 MeV \cite{Wu:2016vtq} (see \cite{An:2022qpt} for review), from below $\eta_c\eta_c$ (5968 MeV) threshold to $\chi_{c1}\chi_{c1}$ threshold (7020 MeV). If the mass lies below di-charmonium threshold, the di-charmonium decay channel is forbidden, two dominant decay channels remain: $J/\psi+X$ or $D\bar{D}$. Under the parameter set of this work, We estimate the width of $T_{4c} \to c\bar{c}$ as reported in Ref.\cite{Zhang:2023ffe}. We find that the $c\bar{c}$ width is slightly smaller than $\gamma J/\psi$ width, not to mention the much lower reconstruction efficiency of $D$-meson. Hence the only possible reconstruct channel is $J/\psi+X$, of which $J/\psi+\gamma$ is the best one. If the $T_{4c}^{0++}$ mass lies above $\eta_c\eta_c$ but below $J/\psi J/\psi$ threshold ($T_{4c}^{0++} \to \eta_c\eta_c \approx 0.1-7$ MeV \cite{liu:2020eha,Xia:2025mgk}), the $\gamma J/\psi$ mode is still the best channel in experiment due to the lower fraction and reconstruction efficiency of $\eta_c$ (e.g., Br[$\eta_c \to KK\pi$] = 7.1\% and the $\eta_c$ is more difficult to trig than $J/\psi$). According to the newly reported measurement released by CMS Collaboration \cite{CMS:2026tiu}, it seems that there are two bumps around 6.2-6.4 GeV in $J/\psi J/\psi$ invariant-mass spectrum. The first bump may caused by $J/\psi J/\psi$ background, the second bump may be attributed to a resonance as predicted by Ref.\cite{Dong:2020nwy}. We note that if there is indeed a $T_{4c}$ state around the $J/\psi J/\psi$ threshold, distinguishing such a state from the dominant background in this mass region remains challenging. Hence a detailed search using $\gamma J/\psi$ invariant-mass spectrum may helpful to clarify this concern.

For $2^{++}$ tetraquark, the dominant decay mode is di-charmonium, the $J/\psi J/\psi$ width is predicted to be 0.36 MeV in the extended relativized quark model \cite{Lu:2020cns}, 3.14 MeV in the compact diquark-antidiquark model \cite{Xia:2025mgk} and 10-80 MeV in covariant quark model \cite{Lu:2025lyu}. The significant difference in the predicted decay widths highlights our still-incomplete understanding of the 2++ tetraquark state. Although its quantum numbers have been experimentally confirmed, further exploration of its decay channels—such as the $\gamma J/\psi$ mode discussed in this work—is essential for a more comprehensive picture. In this paper, the $\gamma J/\psi$ width is calculated to be 0.132 MeV. Although this is significantly smaller than the $J/\psi J/\psi$ width, the branching ratio for the $\gamma J/\psi$ channel remains phenomenologically substantial due to the experimental advantage of requiring one less $J/\psi$ to reconstruct. Several theoretical models predict a $T_{4c}^{2++}$ ground state with a mass below the $J/\psi J/\psi$ threshold. A search for the $\gamma J/\psi$ decay mode could therefore help determine whether X(6500) is the $2^{++}$ ground state.

Currently, there has been no direct experimental search for the vector fully charm tetraquark state $T_{4c}^{1+-}$ at the LHC. In 2023, the Belle Collaboration reported evidence (3.3 $\sigma$) of a near-threshold enhancement in the $J/\psi\eta_c$ invariant mass spectrum \cite{Belle:2023gln}. In contrast to the $J/\psi J/\psi$ background at the LHC, no significant peaking background is expected in this Belle search. In our calculation, the $\gamma\eta_c$ channel for $T_{4c}$ has a width of 1.02 MeV, while the $J/\psi\eta_c$ width lies between 0.45 and 4.31 MeV \cite{liu:2020eha,Xia:2025mgk}. The $\gamma\eta_c$ channel is expected to be more experimentally efficient, as it does not require the reconstruction of a $J/\psi$.

The electromagnetic radiative decay of tetraquarks provides a valuable tool for exploring hadron structure, helping to clarify the ambiguous nature of certain resonances—specifically, whether they arise primarily from compact four-quark structure or are better interpreted as dynamically generated meson–meson molecular states. A compelling case is provided by the X(3872) particle, whose measured radiative decay \cite{LHCb:2024tpv} properties are found to be more consistent with a compact tetraquark configuration than with a loosely bound meson–meson molecular state.

\subsection{$T_{4b}$ decays into bottomonium plus photon}
In 2018, the LHCb Collaboration performed a search for the $T_{4b}$ state using the $\Upsilon\mu^+\mu^-$ invariant-mass spectrum \cite{LHCb:2018uwm}. The analysis utilized data collected by the LHCb detector at center-of-mass energies of $\sqrt{s}$ = 7, 8 and 13 TeV, corresponding to an integrated luminosity of 6.3 $\rm fb^{-1}$. No significant excess was observed. Subsequently, the CMS Collaboration conducted a search using their detector at $\sqrt{s}$ = 13 TeV, with an integrated luminosity of 35.9 $\rm fb^{-1}$ \cite{CMS:2020qwa}. This analysis found only a slight indication of a potential resonance around 19 GeV, with a local significance of approximately $1\sigma$, far below the $5\sigma$ threshold required for discovery. The $T_{4b}$ states share the same quantum numbers as their charm counterparts $T_{4c}$. The S-wave ground states are $0^{++}, 1^{+-}$ and $2^{++}$, with masses estimated to lie between 18 and 20 GeV. The decay width to di-bottomonium is estimated to be of the order of 1 MeV \cite{liu:2020eha,Xia:2025mgk}.
\begin{table}[ht]
	\caption{The decay widths (in unit eV) of $T_{4b} \to \gamma+\Upsilon(\eta_b)$ in this work.}
	\begin{center}
		\centering
		\begin{tabular}{|m{2.cm}<{\centering}|m{2.cm}<{\centering}|m{2.cm}<{\centering}|m{2.cm}<{\centering}|m{2.cm}<{\centering}|}
			\toprule
			\hline
			$\Gamma(\ \rm eV)$ & $0^{++}_{\rm I}$ & $0^{++}_{\rm II}$ & $1^{+-}$ & $2^{++}$ \\
			\hline
			$\gamma \Upsilon$  & 0.975            & 13.3              & -        & 15.2    \\
			\hline 
			$\gamma\eta_b$     & -                & -                 &  39.1    & -        \\
			\hline 		
		\end{tabular}
	\end{center}    
	\label{Tab_Width_T4b}
\end{table}

The analytic expression for the decay width can be readily derived from the corresponding $T_{4c}$ result by substituting $m_c \to m_b$ and $e_c \to e_b$. Using the estimated LDMEs for $T_{4b}$ provided above, we then compute the numerical widths for $T_{4b} \to \gamma\Upsilon(\eta_b)$, which are presented in Table \ref{Tab_Width_T4b}. These widths are approximately 4-6 order of magnitudes smaller than those for $T_{4c}$, lying at the 1-40 eV level. Consequently, the relatively small decay width renders this channel less promising for experimental observation compared to the di-$\Upsilon$ channel. However, it is crucial to note that our current estimates of the LDMEs are based on rather rough approximations. If the actual LDMEs are significantly larger, the decay width for this channel could increase substantially. Ultimately, these possibilities must be tested and constrained by future, more precise experimental data.

\subsection{$T_{bc\bar{b}\bar{c}}$ decays into quarkonium plus photon}
Beyond $T_{4c}$ and $T_{4b}$, only the $T_{bc\bar{b}\bar{c}}$ state is capable of undergoing radiative decay into quarkonium. In contrast, other tetraquarks containing both $b$ and $c$ quarks exhibit different decay patterns: the $T_{bb\bar{c}\bar{c}}$ cannot decay radiatively to a quarkonium state, while the $T_{bb\bar{b}\bar{c}}$ and $T_{cc\bar{c}\bar{b}}$ configurations can only radiate a photon and produce a $B_c$ meson. Here, we calculate the radiative decay widths for the processes $T_{bc\bar{b}\bar{c}} \to \gamma + J/\psi$, $\Upsilon$ and $\eta_{c/b}$. The analytical expressions are provided in Appendix B, for a given quantum number, e.g., $0^{++}$, the physical decay width is a mixture of all possible color-spin configurations. For simplicity, we only calculate the decay width for each configuration individually and do not consider the mixing effect, and the corresponding numerical results are presented in Tables \ref{Tab_Width_Tbcbc2rV} and \ref{Tab_Width_Tbcbc2rPS}.
\begin{table}[htb]
	\centering
	\caption{The widths of $T_{bc\bar{b}\bar{c}}$ decay into vector quarkonium plus a photon. Here ``A'' denotes antisymmetric organization of the color-spin configurations: $\dfrac{1}{\sqrt{2}}\Big[ \ket{(bc)_{\bm{6}}^{1}\mbox{-}(\bar{b}\bar{c})_{\bar{\bm{6}}}^{0}}-\ket{(bc)_{\bm{6}}^{0}\mbox{-}(\bar{b}\bar{c})_{\bar{\bm{6}}}^{1}} \Big]$.}
	\begin{tabular}{|m{1.8cm}<{\centering}|m{1.6cm}<{\centering}|m{1.6cm}<{\centering}|m{1.6cm}<{\centering}|m{1.6cm}<{\centering}|m{1.8cm}<{\centering}|m{1.8cm}<{\centering}|m{1.6cm}<{\centering}|m{1.6cm}<{\centering}|}
		\hline
		\multirow{2}{*}{$\Gamma$} & \multicolumn{4}{c|}{$0^{++}$} & \multicolumn{2}{c|}{$1^{++}$} & \multicolumn{2}{c|}{$2^{++}$} \\
		\cline{2-9}
		 & {\footnotesize $(bc)_{\bm{6}}^{0}\mbox{-}(\bar{b}\bar{c})_{\bar{\bm{6}}}^{0}$} &  {\footnotesize$(bc)_{\bm{6}}^{1}\mbox{-}(\bar{b}\bar{c})_{\bar{\bm{6}}}^{1}$} & {\footnotesize$(bc)_{\bar{\bm{3}}}^{0}\mbox{-}(\bar{b}\bar{c})_{\bm{3}}^{0}$} & {\footnotesize$(bc)_{\bar{\bm{3}}}^{1}\mbox{-}(\bar{b}\bar{c})_{\bm{3}}^{1}$} & {\footnotesize$(bc)_{\bm{6}}^{1}\mbox{-}(\bar{b}\bar{c})_{\bar{\bm{6}}A}^{0}$} & {\footnotesize$(bc)_{\bar{\bm{3}}}^{1}\mbox{-}(\bar{b}\bar{c})_{\bm{3}A}^{0}$} & {\footnotesize$(bc)_{\bm{6}}^{1}\mbox{-}(\bar{b}\bar{c})_{\bar{\bm{6}}}^{1}$} & {\footnotesize$(bc)_{\bar{\bm{3}}}^{1}\mbox{-}(\bar{b}\bar{c})_{\bm{3}}^{1}$}\\
		\hline
		$\gamma J/\psi$ (eV)   & 1.26 & 6.20 & 1.42 & 2.59 & 3.56 & 1.03 & 40.5 & 15.2\\
		\hline
		$\gamma\Upsilon$ (keV) & 28.2  & 148  & 28.7 & 52.3  & 8.54  & 5.13  & 82.0 & 14.5\\
		\hline
	\end{tabular}
    \label{Tab_Width_Tbcbc2rV}
\end{table}

\begin{table}[htb]
	\centering
	\caption{The widths of $T_{bc\bar{b}\bar{c}}$ decay into pseudoscalar quarkonium plus a photon. Here ``S'' denotes the symmetric organization of the color-spin configurations: $\dfrac{1}{\sqrt{2}}\Big[ \ket{(bc)_{\bm{6}}^{1}\mbox{-}(\bar{b}\bar{c})_{\bar{\bm{6}}}^{0}}+\ket{(bc)_{\bm{6}}^{0}\mbox{-}(\bar{b}\bar{c})_{\bar{\bm{6}}}^{1}} \Big]$.}
	\begin{tabular}{|m{1.8cm}<{\centering}|m{1.6cm}<{\centering}|m{1.6cm}<{\centering}|m{1.8cm}<{\centering}|m{1.8cm}<{\centering}|}
		\hline
		\multirow{2}{*}{$\Gamma$} & \multicolumn{4}{c|}{$1^{+-}$} \\
		\cline{2-5}
		& {\footnotesize$(bc)_{\bm{6}}^{1}\mbox{-}(\bar{b}\bar{c})_{\bar{\bm{6}}}^{1}$} &     {\footnotesize$(bc)_{\bar{\bm{3}}}^{1}\mbox{-}(\bar{b}\bar{c})_{\bm{3}}^{1}$}   &  {\footnotesize$(bc)_{\bm{6}}^{1}\mbox{-}(\bar{b}\bar{c})_{\bar{\bm{6}}S}^{0}$}  & {\footnotesize$(bc)_{\bar{\bm{3}}}^{1}\mbox{-}(\bar{b}\bar{c})_{\bm{3}S}^{0}$}  \\
		\hline
		$\gamma\eta_c$ (eV)  & 0.127 & 0.036 & 6.05 & 6.90 \\
		\hline
		$\gamma\eta_b$ (keV) & 47.7   & 19.4  & 133.6 & 66.1 \\
		\hline
	\end{tabular}
	\label{Tab_Width_Tbcbc2rPS}
\end{table}

It is striking to find that the decay width of $T_{bc\bar{b}\bar{c}}$ to $\gamma$+bottomonium is large that to $\gamma$+charmonium by 3-4 orders of magnitude. This significant disparity suggests that the $T_{bc\bar{b}\bar{c}}$ prefers to annihilate its $c\bar{c}$-pair rather than its $b\bar{b}$-pair during the radiative decay process. Nevertheless, we find that the width for $\gamma\Upsilon$ is much smaller than that for $J/\psi\Upsilon$. For comparison, the decay width of the tensor $T_{bc\bar{b}\bar{c}}$ state to $J/\psi\Upsilon$ is reported to be on the order of 27 MeV in Ref. \cite{Agaev:2024qbh}. This implies that the rearrangement decay mode is the dominant process for the $T_{bc\bar{b}\bar{c}}$ tetraquark. 

We also emphasize the crucial role of the LDMEs for $T_{bc\bar{b}\bar{c}}$, as they are essential for reliably estimating the decay widths. Our numerical results provide estimations for the decay widths of individual color-spin configurations. However, the physical decay width should be a coherent superposition of contributions from all configurations, including how to employ interference terms. More accurate determinations of the LDMEs for these individual configurations and their interference terms are therefore needed, if those LDMEs are determined, one can easily to extend to the physical decay width based on the description of this paper.

\section{SUMMARY AND CONCLUSIONS}
In this work, we investigate the fully heavy tetraquark, including $T_{4c}$, $T_{4b}$ and $T_{bc\bar{b}\bar{c}}$, radiatively decay into quarkonium using the NRQCD factorization framework. The analytic squared amplitudes are provided and the LDMEs for $T_{4b}$ and $T_{bc\bar{b}\bar{c}}$ are estimated via non-relativistic Cornell potential. The numerical results underscore the critical role of radiative decay channels in resolving the spectroscopy of exotic tetraquarks. The $J/\psi\gamma$ channel, in particular, offers a promising, low-background signature for experimental searches, especially for states whose masses lie below the di-charmonium threshold. Hence we suggest a experimental research for this channel at the LHC in the future. As the best channel, the measured branching fractions of $T_{4c} \to J/\psi\gamma$ can be used to determine the LDMEs for $T_{4c}$, which are then widely applicable to its production and decay. 

\vspace{1.4cm} {\bf Acknowledgments}
This work is supported by the National Natural Science Foundation of China (NSFC) under the Grants No. 12275185 and No. 12335002.

\appendix
\section{$T_{4c} \to H_{c\bar{c}}+\gamma$}
The analytical squared amplitudes of $T_{4c} \to J/\psi(\eta_c)\gamma$
\begin{align}
	&\big| \mathcal{M}^{0++}_{\bm{6}\otimes\bar{\bm{6}}}(T_{4c} \to J/\psi\gamma) \big|^2 = \dfrac{4620800e_c^2\pi^5\alpha\alpha_s^4}{729m_c^{10}},\nonumber \\
	&\big| \mathcal{M}^{0++}_{\bar{\bm{3}}\otimes\bm{3}}(T_{4c} \to J/\psi\gamma) \big|^2 = \dfrac{24596276224e_c^2\pi^5\alpha\alpha_s^4}{54675m_c^{10}},\nonumber \\
	&\big| \mathcal{M}^{0++}_{\rm \bm{mix}}(T_{4c} \to J/\psi\gamma) \big|^2 = -\dfrac{47676928e_c^2\pi^5\alpha\alpha_s^4}{729m_c^{10}}\sqrt{\dfrac{2}{3}},\nonumber \\
	&\big| \mathcal{M}^{2++}_{\bar{\bm{3}}\otimes\bm{3}}(T_{4c} \to J/\psi\gamma) \big|^2 = \dfrac{11542267904e_c^2\pi^5\alpha\alpha_s^4}{54675m_c^{10}},\nonumber \\
	&\big| \mathcal{M}^{1+-}_{\bar{\bm{3}}\otimes\bm{3}}(T_{4c} \to \eta_c\gamma) \big|^2 = \dfrac{9885304832e_c^2\pi^5\alpha\alpha_s^4}{6075m_c^{10}},
\end{align}
where $e_c$ is the charge of charm quark. The squared amplitudes of $T_{4b} \to \Upsilon(\eta_b)\gamma$ can be easily obtained by replacing $e_c \to e_b$ and $m_c \to m_b$.

\section{$T_{bc\bar{b}\bar{c}} \to H_{b\bar{b}/c\bar{c}}+\gamma$}
The analytical squared amplitudes of $T_{bc\bar{b}\bar{c}}^{0++} \to J/\psi\gamma$
{\footnotesize	
	\begin{align}
		&\big| \mathcal{M}^{0++}_{ \bm{6}^0 \otimes \bar{\bm{6}}^0 }(J/\psi\gamma) \big|^2 = \dfrac{16\pi^5\alpha\alpha_s^4}{6561m_b^{10}}\dfrac{ (48r^7-288r^6+320r^5+685r^4-1217r^3+232r^2+558r-252)^2 }{(3-r)^2(1-r)^5r^2(r^2-4r+2)^2},\nonumber \\
		&\big| \mathcal{M}^{0++}_{ \bm{6}^1 \otimes \bar{\bm{6}}^1 }(J/\psi\gamma) \big|^2 = \dfrac{16\pi^5\alpha\alpha_s^4}{19683m_b^{10}}\dfrac{(48r^7-496r^6+1920r^5-3911r^4+5475r^3-4712r^2+1334r+84)^2}{(3-r)^2(1-r)^5r^2(r^2-4r+2)^2},\nonumber \\
		&\big| \mathcal{M}^{0++}_{ \bar{\bm{3}}^0 \otimes \bm{3}^0 }(J/\psi\gamma) \big|^2 = \dfrac{32\pi^5\alpha\alpha_s^4}{6561m_b^{10}}\dfrac{ (48r^7-384r^6+992r^5-707r^4-593r^3+712r^2+270r-252)^2 }{(3-r)^2(1-r)^5r^2(r^2-4r+2)^2},\nonumber \\
		&\big| \mathcal{M}^{0++}_{ \bar{\bm{3}}^1 \otimes \bm{3}^1 }(J/\psi\gamma) \big|^2 = \dfrac{32\pi^5\alpha\alpha_s^4}{19683m_b^{10}}\dfrac{(48r^7-400r^6+1248r^5-2231r^4+3411r^3-3464r^2+1046r+84)^2}{(3-r)^2(1-r)^5r^2(r^2-4r+2)^2},
	\end{align}
}
where $r = \frac{m_b}{m_b+m_c}$.

The squared amplitudes of $T_{bc\bar{b}\bar{c}}^{1+-} \to \eta_c\gamma$ 
{\footnotesize	
	\begin{align}
		&\big| \mathcal{M}^{1+-}_{ \bm{6}^1 \otimes \bar{\bm{6}}^1 }(\eta_c\gamma) \big|^2 = \dfrac{8\pi^5\alpha\alpha_s^4 }{2187m_b^{10}(3-r)^2(1-r)^7r^2(r^2-4r+2)^2(r^2-3r+1)^2}\nonumber \\ &(-32r^9+453r^8-2674r^7+8544r^6-16257r^5+19240r^4-14268r^3+6436r^2-1604r+168)^2,\nonumber \\
		&\big| \mathcal{M}^{1+-}_{ \bar{\bm{3}}^1 \otimes \bm{3}^1 }(\eta_c\gamma) \big|^2 = \dfrac{16\pi^5\alpha\alpha_s^4}{2187m_b^{10}(3-r)^2(1-r)^7r^2(r^2-4r+2)^2(r^2-3r+1)^2}\nonumber \\ &(-32r^9+405r^8-2242r^7+7104r^6-14001r^5+17464r^4-13596r^3+6340r^2-1604r+168)^2,\nonumber \\
		&\big| \mathcal{M}^{1+-}_{ \bm{6}^1 \otimes \bar{\bm{6}}^0[S]}(\eta_c\gamma) \big|^2 = \dfrac{8\pi^5\alpha\alpha_s^4}{2187m_b^{10}(3-r)^2(1-r)^7r^2(r^2-4r+2)^2(r^2-3r+1)^2}\nonumber \\ &(-32r^9+95r^8+1322r^7-9124r^6+23305r^5-28120r^4+14556r^3-340r^2-2172r+504)^2,\nonumber \\
		&\big| \mathcal{M}^{1+-}_{ \bar{\bm{3}}^1 \otimes \bm{3}^0[S]}(\eta_c\gamma) \big|^2 = \dfrac{16\pi^5\alpha\alpha_s^4}{2187m_b^{10}(3-r)^2(1-r)^7r^2(r^2-4r+2)^2(r^2-3r+1)^2}\nonumber \\ &(-32r^9+239r^8-166r^7-3268r^6+12313r^5-17992r^4+10236r^3+332r^2-2172r+504)^2.
	\end{align}
}

The squared amplitudes of $T_{bc\bar{b}\bar{c}}^{1++} \to J/\psi\gamma$
{\footnotesize	
	\begin{align}
		&\big| \mathcal{M}^{1++}_{ \bm{6}^1 \otimes \bar{\bm{6}}^0[A]}(J/\psi\gamma) \big|^2 = \dfrac{512\pi^5\alpha\alpha_s^4}{6561m_b^{10} (3-r)^2(1-r)^5r^2(r^2-4r+2)^2}(169r^{12}-2444r^{11}+14755r^{10}-49502r^9 \nonumber \\
		&\hspace{0.2cm}+107850r^8-178666r^7+248618r^6-276692r^5+229705r^4-140004r^3+64916r^2-21504r+3528) \nonumber \\
		&\big| \mathcal{M}^{1++}_{ \bar{\bm{3}}^1 \otimes \bm{3}^0[A]}(J/\psi\gamma) \big|^2 = \dfrac{1024\pi^5\alpha\alpha_s^4}{6561m_b^{10} (3-r)^2(1-r)^5r^2(r^2-4r+2)^2}(r^{12}+4r^{11}-53r^{10}-182r^9+2370r^8 \nonumber \\
		&\hspace{1cm}-6454r^7+3182r^6+13660r^5-22007r^4+1788r^3+19844r^2-15456r+3528).
	\end{align}
}

Here ``A'' and ``S'' denote the antisymmetric and symmetric organizations of the color-spin structures, e.g., $\dfrac{1}{\sqrt{2}}\Big[ \ket{(bc)_{\bm{6}}^{1}\mbox{-}(\bar{b}\bar{c})_{\bar{\bm{6}}}^{0}}\mp\ket{(bc)_{\bm{6}}^{0}\mbox{-}(\bar{b}\bar{c})_{\bar{\bm{6}}}^{1}} \Big]$. 

The squared amplitudes of $T_{bc\bar{b}\bar{c}}^{2++} \to J/\psi\gamma$ 
{\footnotesize
	\begin{align}
		&\big| \mathcal{M}^{2++}_{ \bm{6}^1 \otimes \bar{\bm{6}}^1 }(J/\psi\gamma) \big|^2 = \dfrac{512\pi^5\alpha\alpha_s^4}{19683m_b^{10} (3-r)^2(1-r)^5r^2(r^2-4r+2)^2}(676r^{12}-11856r^{11}+91984r^{10}-415008r^9 \nonumber \\
		&\hspace{0.2cm}+1201580r^8-2315464r^7+2947340r^6-2289776r^5+767335r^4+216004r^3-196724r^2-6720r+17640) \nonumber \\
		&\big| \mathcal{M}^{2++}_{ \bar{\bm{3}}^1 \otimes \bm{3}^1 }(J/\psi\gamma) \big|^2 = \dfrac{1024\pi^5\alpha\alpha_s^4}{19683m_b^{10} (3-r)^2(1-r)^5r^2(r^2-4r+2)^2}(196r^{12}-3024r^{11}+20812r^{10}-84336r^9 \nonumber \\
		&\hspace{0.2cm}+222176r^8-393724r^7+456392r^6-270320r^5-82985r^4+240580r^3-80228r^2-36960r+17640).
	\end{align}
}

The analytical squared amplitudes of $T_{bc\bar{b}\bar{c}}^{0++} \to \Upsilon\gamma$
{\footnotesize	
	\begin{align}
		&\big| \mathcal{M}^{0++}_{ \bm{6}^0 \otimes \bar{\bm{6}}^0 }(\Upsilon\gamma) \big|^2 = \dfrac{64\pi^5\alpha\alpha_s^4}{6561m_b^{10}}\dfrac{ r^5(12r^7-51r^6-166r^5+259r^4+250r^3-414r^2+322r-86)^2 }{(1-r)^{12}(r+2)^2(r^2+2r-1)^2},\nonumber \\
		&\big| \mathcal{M}^{0++}_{ \bm{6}^1 \otimes \bar{\bm{6}}^1 }(\Upsilon\gamma) \big|^2 = \dfrac{64\pi^5\alpha\alpha_s^4}{19683m_b^{10}}\dfrac{r^5(12r^7+r^6+30r^5+599r^4+398r^3-1318r^2-190r+258)^2}{(1-r)^{12}(r+2)^2(r^2+2r-1)^2},\nonumber \\
		&\big| \mathcal{M}^{0++}_{ \bar{\bm{3}}^0 \otimes \bm{3}^0 }(\Upsilon\gamma) \big|^2 = \dfrac{128\pi^5\alpha\alpha_s^4}{6561m_b^{10}}\dfrac{ r^5(12r^7+9r^6-106r^5-125r^4+106r^3+198r^2+118r-86)^2 }{(1-r)^{12}(r+2)^2(r^2+2r-1)^2},\nonumber \\
		&\big| \mathcal{M}^{0++}_{ \bar{\bm{3}}^1 \otimes \bm{3}^1 }(\Upsilon\gamma) \big|^2 = \dfrac{128\pi^5\alpha\alpha_s^4}{19683m_b^{10}}\dfrac{r^5(12r^7+13r^6+42r^5+407r^4+254r^3-850r^2-346r+258)^2}{(1-r)^{12}(r+2)^2(r^2+2r-1)^2}.
	\end{align}
}

The squared amplitudes of $T_{bc\bar{b}\bar{c}}^{1+-} \to \eta_b\gamma$
{\footnotesize	
	\begin{align}
		&\big| \mathcal{M}^{1+-}_{ \bm{6}^1 \otimes \bar{\bm{6}}^1 }(\eta_b\gamma) \big|^2 = \dfrac{8\pi^5\alpha\alpha_s^4r^3 (16r^9+117r^8+404r^7+559r^6-468r^5-1709r^4+32r^3+1325r^2-456r+12)^2}{2187m_b^{10}(1-r)^{12}(r+2)^2(r^2+r-1)^2(r^2+2r-1)^2}\nonumber \\ 
		&\big| \mathcal{M}^{1+-}_{ \bar{\bm{3}}^1 \otimes \bm{3}^1 }(\eta_b\gamma) \big|^2 = \dfrac{16\pi^5\alpha\alpha_s^4r^3 (16r^9+165r^8+596r^7+463r^6-1188r^5-1373r^4+800r^3+653r^2-312r+12)^2}{2187m_b^{10}(1-r)^{12}(r+2)^2(r^2+r-1)^2(r^2+2r-1)^2}\nonumber \\ 
		&\big| \mathcal{M}^{1+-}_{ \bm{6}^1 \otimes \bar{\bm{6}}^0 }(\eta_b\gamma) \big|^2 = \dfrac{8\pi^5\alpha\alpha_s^4r^3 (16r^9-287r^8-1236r^7-129r^6+3128r^5+935r^4-2256r^3+133r^2+212r-12)^2}{2187m_b^{10}(1-r)^{12}(r+2)^2(r^2+r-1)^2(r^2+2r-1)^2}\nonumber \\
		&\big| \mathcal{M}^{1+-}_{ \bar{\bm{3}}^1 \otimes \bm{3}^0 }(\eta_b\gamma) \big|^2 = \dfrac{16\pi^5\alpha\alpha_s^4r^3 (16r^9-143r^8-852r^7-609r^6+1928r^5+1943r^4-1296r^3-932r^2+452r-12)^2}{2187m_b^{10}(1-r)^{12}(r+2)^2(r^2+r-1)^2(r^2+2r-1)^2}.
	\end{align}
}

The squared amplitudes of $T_{bc\bar{b}\bar{c}}^{1++} \to \Upsilon\gamma$
{\footnotesize	
	\begin{align}
		&\big| \mathcal{M}^{1++}_{ \bm{6}^1 \otimes \bar{\bm{6}}^0 }(\Upsilon\gamma) \big|^2 = \dfrac{32\pi^5\alpha\alpha_s^4r^5}{6561m_b^{10} (1-r)^{12}(r+2)^2(r^2+2r-1)^2}(10816r^{12}+39104r^{11}-31239r^{10}
		-112744r^9\nonumber \\
		&+287484r^8+109864r^7-886058r^6+491792r^5+729916r^4-1074104r^3+591049r^2-159192r+17424) \nonumber \\
		&\big| \mathcal{M}^{1++}_{ \bar{\bm{3}}^1 \otimes \bm{3}^0 }(\Upsilon\gamma) \big|^2 = \dfrac{64\pi^5\alpha\alpha_s^4r^5}{6561m_b^{10} (1-r)^{12}(r+2)^2(r^2+2r-1)^2}(64r^{12}-64r^{11}-615r^{10}+1304r^9+2940r^8 \nonumber \\
		&\hspace{1cm}-6296r^7-4202r^6+18128r^5+7612r^4-6392r^3+5737r^2-5400r+1296).
	\end{align} 
}

The squared amplitudes of $T_{bc\bar{b}\bar{c}}^{2++} \to \Upsilon\gamma$
{\footnotesize
	\begin{align}
		&\big| \mathcal{M}^{2++}_{ \bm{6}^1 \otimes \bar{\bm{6}}^1 }(\Upsilon\gamma) \big|^2 = \dfrac{32\pi^5\alpha\alpha_s^4r^5}{19683m_b^{10} (1-r)^{12}(r+2)^2(r^2+2r-1)^2}(16900r^{12}+118560r^{11}+263923r^{10}+62128r^9 \nonumber \\
		&\hspace{0.2cm}-518380r^8-599736r^7-100758r^6+240064r^5+550752r^4+369608r^3-239285r^2-155424r+62208) \nonumber \\
		&\big| \mathcal{M}^{2++}_{ \bar{\bm{3}}^1 \otimes \bm{3}^1 }(\Upsilon\gamma) \big|^2 = \dfrac{64\pi^5\alpha\alpha_s^4r^5}{19683m_b^{10} (1-r)^{12}(r+2)^2(r^2+2r-1)^2}(1156r^{12}+4896r^{11}+5395r^{10}-4304r^9-18796r^8 \nonumber \\
		&\hspace{0.2cm}-20472r^7-39126r^6-60416r^5+123744r^4+189704r^3-62357r^2-74208r+25344).
	\end{align}
}

\end{document}